\documentclass{report}
\usepackage{amssymb}
\usepackage{amsfonts}
\usepackage{amsmath}

\input{tcilatex}

\begin{document}

\title{The Problem of Modelling of Economic Dynamics in Differential Form}
\author{S. I. Chernyshov \and V. S. Ponomarenko  \and A. V. Voronin}
\date{}
\maketitle
\tableofcontents

\bigskip \newpage

\bigskip

\begin{center}
{\LARGE Introduction}
\end{center}

The authors have tried to analyze procedures of the construction of
differential equations that are employed for modeling of macroeconomic
processes. The results prove to be rather unexpected. Thus, the derivation
of the differential equation of Harrod's model is based on a linear relation
between capital and income. As a result, there arises a contradiction in
terms of dimension that is rooted in incorrect treatment of the fundamental
notion of the infinitesimal quantity. One can overcome this contradiction by
relating capital to the integral of income over a corresponding time
interval. However, in this case, the solution is by no means an exponential
growth but a much more realistic relation that reflects, in particular,
objective finiteness of the prognostic period.

An analysis of the models of Harrod-Domar, Phillips, as well as of other
models (see the well-known treatise by R. Allen), leads us to the conclusion
that analogous deficiencies are, in principle, inherent in these models too.
In general, the refraction in the sphere of economic dynamics of the
methodology of the construction of mathematical models borrowed from the
field of natural sciences, such as dynamics, electrodynamics, etc., proves
to be absolutely unjustified. As a matter of fact, differential equations
adequate to these models follow naturally from the consideration of an
infinitesimal element. However, as regards the problems of economics, such
an approach is objectively senseless. Nevertheless, economics, in its turn,
has intrinsic advantages from the point of view of possibilities of
mathematical modeling, which is embodied in the notion of balance. As we
will show, there exist formal means to reduce Leontief's model of
"expenses-output" in its canonical interpretation to a system of linear
differential equation (of, generally speaking, arbitrary order with respect
to the derivatives).

At the same time, the scantiness of the arsenal of the means of linear
theory that are used in representative modeling of macroeconomic processes
is almost universally recognized nowadays. In this regard, we will
characterize briefly those areas of systems analysis that are devoted to the
construction of nonlinear models that are adequate to a given "input-output"
mapping. In what follows, we nonetheless note that Leontief's model in the
differential form can be elementary reduced to a Fredholm integral equation
of the second kind (with respect to a vector function), whose theory and
algorithms of numerical realization are as constructive as possible. In the
case, when the kernel of such an equation depends on a parameter, which is
quite naturally interpreted in terms of the object sphere, the spectrum of
its possible solutions becomes extremely wide. We think that the development
of the theory of Fredholm integral equations of the second kind, whose
kernels contain parameters, and its application to the modeling of the
processes of economic dynamics is rather promising.

\textit{Note on literature references in the English version:} The reader
should be advised that all the references to page and section numbers
appearing in the text are given according to the Russian editions of
corresponding literature sources.

\textit{Contact address}: voronin@hneu.edu.ua

\chapter{Analysis of Harrod's model}

\section{Computational relations}

Let us turn to Harrod's model of the development of the economy represented
in the book by L. V. Kantorovich and A. B. Gorstko \cite{1} (pp.160, 161).
In the authors' view, "despite the simplicity - the model takes into account
only one limited factor, i.e., the capital - it can be used for a rude
approximate investigation into the laws of the growth of the economy". In
the latter context, we focus our attention on the above-mentioned
exclusiveness of the capital. The employed notation are explained as
follows: $Y\left( t\right) $ is the national income; $K\left( t\right) $ is
the capital (the assets); $C\left( t\right) $ is the volume of consumption; $%
S\left( t\right) $ is the volume of accumulation, and $I\left( t\right) $ is
the investment. All the quantities are dimensional. For the analysis of the
model, it is reasonable to reproduce the text \cite{1}.

"It is assumed that the economy is functioning in such a way that the
following relations are fulfilled:%
\begin{equation}
Y\left( t\right) =C\left( t\right) +S\left( t\right)  \label{1.1.1}
\end{equation}%
(the national income is distributed between accumulation and consumption); 
\begin{equation}
S\left( t\right) =I\left( t\right)  \label{1.1.2}
\end{equation}%
(the accumulation is equal to the investment);%
\begin{equation}
S\left( t\right) =\mu Y\left( t\right) ,  \label{1.1.3}
\end{equation}%
where $\mu =const$ (the accumulation constitutes a constant fraction of the
national income);%
\begin{equation}
\dot{K}\left( t\right) =I\left( t\right)  \label{1.1.4}
\end{equation}%
(the growth rate of the capital is equal to the investment, where $\dot{K}%
=dK/dt$);%
\begin{equation}
K\left( t\right) =\nu Y\left( t\right) ,  \label{1.1.5}
\end{equation}%
where $\nu =const$ (the capital-to-national income ratio is a constant
quantity, which is an observed empirical fact).

From the equations that describe the model, it follows immediately:%
\begin{equation}
Y\left( t\right) =\frac{1}{\mu }I\left( t\right) =\frac{1}{\mu }\dot{K}%
\left( t\right) =\frac{\nu }{\mu }\dot{Y}\left( t\right)  \label{1.1.6}
\end{equation}%
or%
\begin{equation}
\dot{Y}\left( t\right) /Y\left( t\right) =\mu /\nu .  \label{1.1.7}
\end{equation}

Consequently, the growth rate of the national income is equal to $\mu /\nu $%
, and, therefore, if one assumes that at the initial moment of time (at $t=0$%
) the national income is equal to $Y_{0}$, the law of its change with time
has the following form:%
\begin{equation}
Y\left( t\right) =Y_{0}e^{\mu t/\nu },  \label{1.1.8}
\end{equation}%
which usually stands in quite good agreement with practice".

In obtaining relations (\ref{1.1.6}) and differential equation (\ref{1.1.7}%
), a key role is played by the dependence%
\begin{equation}
K\left( t\right) =\int\limits_{-T}^{t}I\left( \eta \right) d\eta
\label{1.1.9}
\end{equation}%
that is obvious. Here, the capital $K\left( t\right) $ can be measured only
monetary equivalent whose unit will be chosen to be the dollar ($\$$); $t$
is the time variable measured, for definiteness, in seconds ($s$). Given
that $dt$ has the same dimension as $t$, the function $I\left( t\right) $
represents the intensity of the investment flow measured as $\$/s$. A
corollary of (\ref{1.1.9}) is just the derivative (\ref{1.1.4}): there is
simply no other derivative in the model (\ref{1.1.1})-(\ref{1.1.5}).

On the basis of (\ref{1.1.1})-(\ref{1.1.3}), we can conclude that $Y\left(
t\right) $, $C\left( t\right) $ and $S\left( t\right) $ are also intensities
of the flows of the income, consumption, and accumulation ($\$/s$),
respectively. However, there arises a contradiction, because $Y\left(
t\right) $ in (\ref{1.1.5}) is the national income per one-year period
measured in $\$$. Analogously, $t$ in (\ref{1.1.8}) denotes the number of
years (their fractions). Therefore, the exponent $\mu t/\nu $ is
dimensionless. Note that we had to carry out an investigation into the
dimension of the components of the model. For the reason that will be
explained in what follows, one does not specify clearly the dimension of
employed quantities in macroeconomics of mathematical orientation. However,
this issue underlies the most important point of the whole consideration!

\section{Refinement on the content of the model}

As $Y\left( t\right) $ is the intensity of the income, relation (\ref{1.1.5}%
) had to look as follows%
\begin{equation}
K\left( t\right) =\nu \int\limits_{t_{i}}^{t_{i}+t_{\ast }}Y\left( \eta
\right) d\eta ,\quad t\in \left[ t_{i},t_{i}+t_{\ast }\right] ,
\label{1.2.1}
\end{equation}%
where $t_{\ast }$ is the time interval of one year; $\nu $ is the number of
years during which such an income counterbalances the capital of the $i$-th
year.

From a mathematical point of view, such a specification is of no great
importance: what is actually important is the fact that $t_{\ast }$ is
finite. Behind this fact, one can easily see a transition to discrete
analysis and the impossibility of deriving any differential equation! This
major objective of the present investigation will be developed below.

At the same time, one can express the capital $K\left( t\right) $ via the
intensity of the income $Y\left( t\right) $ in a way that is different from
what is done in (\ref{1.2.1}): namely, based on (\ref{1.1.2}) and (\ref%
{1.1.3}), one can employ an analog of (\ref{1.1.9}). Consider relation (\ref%
{1.1.5}) strictly adhering to the definition \cite{1}: "$\nu =const$ (the
capital-to-national income ratio is a constant quantity, which is an
observed empirical fact)". In this regard, let us draw attention to the
informativeness of the coefficient $\nu $ (in contrast, $\mu $ is purely
primitive). The essence is that the "observed fact" characterizes a basic
coefficient $\nu _{\ast }$ for the time interval $t_{\ast }$ during which
the income is measured. For $t_{\ast }=1$ year it is reasonable to set $\nu
_{\ast }=10$. Accordingly, for a certain time interval $\Delta t$, the
coefficient is given by%
\begin{equation}
\nu =\nu _{\ast }t_{\ast }/\Delta t.  \label{1.2.2}
\end{equation}

At that, the flow $Y\left( t\right) $ ($\$/s$) creates, in the period of
time $\Delta t$, the volume of the income ($\$$) equal to%
\begin{equation}
\int\limits_{t}^{t+\Delta t}Y\left( \eta \right) d\eta  \label{1.2.3}
\end{equation}%
that, by the definition \cite{1}, is proportional to the capital $K\left(
t+\Delta t\right) $ with the coefficient $\nu $: see (\ref{1.2.2}).
Accordingly,%
\begin{equation*}
K\left( t+\Delta t\right) =\nu \int\limits_{t}^{t+\Delta t}Y\left( \eta
\right) d\eta ;\quad K\left( t+2\Delta t\right) =\nu
\int\limits_{t}^{t+2\Delta t}Y\left( \eta \right) d\eta .
\end{equation*}%
After differentiation,%
\begin{equation*}
\dot{K}\left( t+\Delta t\right) =\nu \left[ Y\left( t+\Delta t\right)
-Y\left( t\right) \right] ;\quad \dot{K}\left( t+2\Delta t\right) =\nu \left[
Y\left( t+2\Delta t\right) -Y\left( t\right) \right] ,
\end{equation*}%
and, as a result of subtraction,%
\begin{equation*}
\dot{K}\left( t+2\Delta t\right) -\dot{K}\left( t+\Delta t\right) =\nu \left[
Y\left( t+2\Delta t\right) -Y\left( t+\Delta t\right) \right] .
\end{equation*}%
Assuming that the interval $\Delta t$ is sufficiently small, we use a Taylor
expansion:%
\begin{equation*}
\dot{K}\left( t+\Delta t\right) =\dot{K}\left( t\right) +\ddot{K}\left(
t\right) \Delta t+\ldots ;\quad \dot{Y}\left( t+\Delta t\right) =\dot{Y}%
\left( t\right) +\ddot{Y}\left( t\right) \Delta t+\ldots ,
\end{equation*}%
where we have preserved the terms containing the first power of $\Delta t$
and $2\Delta t$. We get:%
\begin{equation}
\ddot{K}\left( t\right) =\nu \dot{Y}\left( t\right) .  \label{1.2.4}
\end{equation}

Thus, the considered model is reduced to the set of relations (\ref{1.1.1})-(%
\ref{1.1.4}) and (\ref{1.2.4}). From these relations, it follows%
\begin{equation*}
\ddot{K}\left( t\right) =\mu \dot{Y}\left( t\right) =\nu \dot{Y}\left(
t\right) ,
\end{equation*}%
which just yields $\mu =\nu $, or, in the case $\mu \neq \nu $,%
\begin{equation*}
Y\left( t\right) =c/\left( \mu -\nu \right) ,
\end{equation*}%
where $c$ is a constant. As regards the derivative $\dot{Y}\left( t\right) $
that, as in (\ref{1.1.6}), leads to a differential equation, it has
treacherously escaped. In general, the solution to the considered problem by
means of (\ref{1.2.3}) can only be trivial.

In other words, in order to equate the function $Y\left( t\right) $ (with
the coefficient) to its derivative in (\ref{1.1.7}) obtaining the
exponential growth (\ref{1.1.8}), it was absolutely necessary to resort to
the surrogate (\ref{1.1.5}) instead of (\ref{1.2.1}). However, the
"derivative" can be understood. Having used (\ref{1.2.3}), we have
essentially obtained an analog of the interrelation between the capital and
the flow (\ref{1.1.9}) without complementing the model at the level of a
quantitative content. These arguments are of heuristic nature.

\section{Improvement of the initial model}

Let us act in a different way, namely, by considering the time interval from 
$0$ to $t$. A priori, we assume that $\Delta t$ is small. The income during
this period is equal to%
\begin{equation*}
\int\limits_{0}^{t}Y\left( \eta \right) d\eta ,
\end{equation*}%
and, by analogy with (\ref{1.2.2}),%
\begin{equation}
\nu =\nu _{\ast }t_{\ast }/t;  \label{1.3.1}
\end{equation}%
however, the assumption that $\nu =const$ in relation (\ref{1.1.5}) should
be rejected. Indeed, it contains a contradiction resulting from the fact
that $\nu $ is related to a year \cite{1} [whose duration $t_{\ast }$ is
related to (\ref{1.2.1})]. In order to remain within the framework of
continuous analysis, we just use the rule of proportion. Thus,%
\begin{equation}
K\left( t\right) =\nu \int\limits_{0}^{t}Y\left( \eta \right) d\eta =\frac{%
\nu _{\ast }t_{\ast }}{t}\int\limits_{0}^{t}Y\left( \eta \right) d\eta .
\label{1.3.2}
\end{equation}

By the way, for $t\rightarrow 0$, the coefficient $\nu \rightarrow \infty $,
which is quite reasonable. Simultaneously, equation (\ref{1.3.2}) takes the
form $K\left( 0\right) =\nu _{\ast }t_{\ast }Y\left( 0\right) $, and it can
be easily compared to (\ref{1.1.5}) and (\ref{1.2.1}) with regard to
dimension. From (\ref{1.3.2}), it follows that%
\begin{equation}
\dot{K}\left( t\right) =-\frac{\nu _{\ast }t_{\ast }}{t^{2}}%
\int\limits_{0}^{t}Y\left( \eta \right) d\eta +\frac{\nu _{\ast }t_{\ast }}{t%
}Y\left( t\right) ,  \label{1.3.3}
\end{equation}%
and, obviously, $\dot{K}\left( t\right) \rightarrow \infty $ for $%
t\rightarrow 0$. By (\ref{1.1.4}), $I\left( t\right) \rightarrow \infty $ as
well. This peculiarity will be eliminated in what follows by the
cancellation of $t$.

From relations (\ref{1.1.2})-(\ref{1.1.4}), we get%
\begin{equation*}
\dot{K}\left( t\right) =\mu Y\left( t\right) ,
\end{equation*}%
which, combined with (\ref{1.3.3}), yields the equation%
\begin{equation*}
\left( 1-\frac{\mu t}{\nu _{\ast }t_{\ast }}\right) t\dot{Y}\left( t\right) =%
\frac{2\mu }{\nu _{\ast }t_{\ast }}tY\left( t\right) ,\quad t>0.
\end{equation*}%
Making a change of the variable%
\begin{equation}
\hat{t}=t/t_{\ast },  \label{1.3.4}
\end{equation}%
by 
\begin{equation*}
\frac{dY\left( t\right) }{dt}=\frac{dY\left( \hat{t}\right) }{d\hat{t}}\frac{%
d\hat{t}}{dt}=\dot{Y}\left( \hat{t}\right) /t_{\ast },
\end{equation*}%
we get%
\begin{equation}
\dot{Y}\left( \hat{t}\right) -\frac{2\sigma }{1-\sigma \hat{t}}Y\left( \hat{t%
}\right) =0,\quad \sigma =\frac{\mu }{\nu _{\ast }}.  \label{1.3.5}
\end{equation}

The solution to this equation has the form%
\begin{equation}
Y\left( \hat{t}\right) =\frac{Y_{0}}{\left( 1-\sigma \hat{t}\right) ^{2}}.
\label{1.3.6}
\end{equation}%
On this basis, the following conclusions can be drawn:

- if $\mu =0$ (the absence of investment) or $\nu _{\ast }\rightarrow \infty 
$ in (\ref{1.3.5}), then $Y\left( \hat{t}\right) =Y_{0}$, which is not
unreasonable;

- the time interval of a reasonable forecast is reflected, because for $\hat{%
t}=\sigma ^{-1}=\nu _{\ast }/\mu $ the income function (\ref{1.3.6}) becomes
senseless;

- generally speaking, such a forecast is inherent in a reliable model, which
should be contrasted with an infinite growth of the function (\ref{1.1.8});

- under the interpretation that $\mu =0.5$ (consumption and accumulation
equally share the income) and $\nu _{\ast }=10$ years, the period of a
conditionally reliable forecast is also equal to 10 years, if it is set
equal to $0.5\sigma ^{-1}$. This seems to be realistic.

It should be emphasized that $Y_{0}$ in (\ref{1.3.6}) is the intensity of
the income for $t=0$ ($\$/s$) rather than an income per year as in (\ref%
{1.1.8}). Note that the coefficient $\nu $ (as the number of years) is
interpreted in sections 1.1 and 1.2, 1.3 as dimensional and dimensionless,
respectively. This fact is a consequence of the observed in section 1.1
contradiction between the interpretation of $Y\left( t\right) $ as the
intensity of the income, according to (\ref{1.1.1})-(\ref{1.1.3}), and the
volume of the income, according to (\ref{1.1.5}).

Thus, based on (\ref{1.3.1}), we have obtained a rather satisfactory result.
It is stipulated by qualitative different dependence of the capital $K\left(
t\right) $ on the flows $I\left( t\right) $ and $Y\left( t\right) $: see (%
\ref{1.1.9}) and (\ref{1.3.2}), respectively. Namely, there is a variable
coefficient $t^{-1}$ in (\ref{1.3.2}).

\section{Discrete essence of the initial model}

Under the interpretation of $Y\left( t\right) $, $C\left( t\right) $, $%
S\left( t\right) $, and $I\left( t\right) $ as flows ($\$/s$), the model is
represented by relations (\ref{1.1.1})-(\ref{1.1.4}) and (\ref{1.2.1}).
Using the definite integral in its simplest interpretation, we can represent
(\ref{1.2.1}) as follows:%
\begin{equation}
K\left( t\right) =\nu \int\limits_{t_{i}}^{t_{i}+t_{\ast }}Y\left( \eta
\right) d\eta =\nu t_{\ast }Y\left( t_{i}\right) ,\quad t\in \left[
t_{i},t_{i}+t_{\ast }\right] ,\quad i=0,1,\ldots .  \label{1.4.1}
\end{equation}

A change of the variables, by (\ref{1.3.4}), leads to the following
relations:%
\begin{equation}
Y\left( \hat{t}\right) =C\left( \hat{t}\right) +S\left( \hat{t}\right)
;\quad S\left( \hat{t}\right) =I\left( \hat{t}\right) ;\quad S\left( \hat{t}%
\right) =\mu Y\left( \hat{t}\right) ;  \label{1.4.2}
\end{equation}%
\begin{equation}
\dot{K}\left( \hat{t}\right) =t_{\ast }I\left( \hat{t}\right) ;
\label{1.4.3}
\end{equation}%
\begin{equation}
K\left( \hat{t}\right) =\nu t_{\ast }Y\left( \hat{t}_{i}\right) ;\quad t\in %
\left[ \hat{t}_{i},\hat{t}_{i}+1\right] ,\quad \hat{t}_{i}=0,1,\ldots ,
\label{1.4.4}
\end{equation}%
where $Y\left( \hat{t}\right) $, $C\left( \hat{t}\right) $, $S\left( \hat{t}%
\right) $, and $I\left( \hat{t}\right) $ are intensities of the flows
measured in $\$/s$.

There is an obvious contradiction related to the fact that the functions in (%
\ref{1.4.2})-(\ref{1.4.4}) cannot, on the one hand, be discrete and, on the
other hand, represent intensities of continuous flows. However, it is
impossible to satisfy (\ref{1.4.4}) continuously, within the framework of
the model (\ref{1.1.1})-(\ref{1.1.5}): see section 1.2. By analogy with (\ref%
{1.4.4}), one is only left with the option to represent relations (\ref%
{1.4.2}) and (\ref{1.4.3}) in a discrete form:%
\begin{equation}
\tilde{Y}\left( \hat{t}_{i}\right) =\tilde{C}\left( \hat{t}_{i}\right) +%
\tilde{S}\left( \hat{t}_{i}\right) ;\quad \tilde{S}\left( \hat{t}_{i}\right)
=\tilde{I}\left( \hat{t}_{i}\right) ;\quad \tilde{S}\left( \hat{t}%
_{i}\right) =\mu \tilde{Y}\left( \hat{t}_{i}\right) ;  \label{1.4.5}
\end{equation}%
\begin{equation}
\dot{K}\left( \hat{t}_{i}\right) =\tilde{I}\left( \hat{t}_{i}\right) ,
\label{1.4.6}
\end{equation}%
where%
\begin{equation}
\tilde{Y}\left( \hat{t}_{i}\right) =t_{\ast }Y\left( \hat{t}_{i}\right) ;\ 
\tilde{C}\left( \hat{t}_{i}\right) =t_{\ast }C\left( \hat{t}_{i}\right) ;\ 
\tilde{S}\left( \hat{t}_{i}\right) =t_{\ast }S\left( \hat{t}_{i}\right) ;\ 
\tilde{I}\left( \hat{t}_{i}\right) =t_{\ast }I\left( \hat{t}_{i}\right) ,
\label{1.4.7}
\end{equation}%
\begin{equation*}
i=0,1,\ldots ,n
\end{equation*}%
have the dimension of capital, which is quite unambiguously said in \cite{1}%
: see section 1.1.

At the same time, under inaccurate treatment, such an approach stands in
disagreement with the rules of the calculus of infinitesimal that form the
basis of differential models. Suppose that investments in years $\hat{t}_{i}$
and $\hat{t}_{i}+1$ are given by $I_{i}$ and $I_{i+1}$, respectively. Then,
the capital is%
\begin{equation*}
K_{i}=K_{i-1}+I_{i};\quad K_{i+1}=K_{i-1}+I_{i}+I_{i+1}.
\end{equation*}%
However, the definition $\dot{K}\left( \hat{t}_{i}\right) =I_{i+1}-I_{i}$ in
discrete analysis is illegitimate.

Should the techniques of continuous analysis be formally extended to a
transformation of relations (\ref{1.4.4})-(\ref{1.4.6}), then, as it is said
about (\ref{1.1.6}) (see section 1.1), it "follows immediately" that 
\begin{equation}
\frac{d\tilde{Y}}{d\hat{t}}\left( \hat{t}_{i}\right) /\tilde{Y}\left( \hat{t}%
_{i}\right) =\mu /\nu ,  \label{1.4.8}
\end{equation}%
and, accordingly,%
\begin{equation}
\tilde{Y}\left( \hat{t}_{i}\right) =\tilde{Y}_{0}e^{\mu \hat{t}_{i}/\nu },
\label{1.4.9}
\end{equation}%
which should be exactly the interpretation of the solution (\ref{1.1.8}).

Thus, all the functions in (\ref{1.4.4})-(\ref{1.4.6}) are discontinuous at $%
\hat{t}=\hat{t}_{i}$. Otherwise, the system of these relations is
degenerate, and the solution is trivial. As a matter of fact, discreteness
stipulates a combination of factors: $Y\left( t\right) \neq Y_{0}$; the
interval $t_{\ast }$ in (\ref{1.4.1}) is finite.

The function $K\left( \hat{t}\right) $ is also discontinuous, because, by (%
\ref{1.1.9}), (\ref{1.3.4}) and (\ref{1.4.7}),%
\begin{equation}
K\left( \hat{t}\right) =\int\limits_{-T/t_{\ast }}^{\hat{t}}\hat{I}\left(
\eta \right) d\eta =K_{0}+\sum_{i=0}^{n}\hat{I}\left( \hat{t}_{i}\right)
,\quad 0\leq \hat{t}<\hat{t}_{n+1},  \label{1.4.10}
\end{equation}%
where%
\begin{equation*}
K_{0}=\int\limits_{-T/t_{\ast }}^{0}\hat{I}\left( \eta \right) d\eta .
\end{equation*}%
As a consequence, this function is not differentiable in a usual sense at $%
\hat{t}=\hat{t}_{i}$.

However, by the existence of (\ref{1.4.9}), the discontinuities of $K\left( 
\hat{t}\right) $ at $\hat{t}=\hat{t}_{i}$ in (\ref{1.4.10}) are finite.
Namely, they are analogous to discontinuities of the function $Y\left( \hat{t%
}\right) $ with the coefficient $\nu $. Summarizing, we arrive at the
conclusion that the function (\ref{1.4.10}) can be understood only as a
generalized function, and, accordingly \cite{2} (pp. 57-60),%
\begin{equation}
\dot{K}\left( \hat{t}\right) =\sum_{i=0}^{n}\hat{I}\left( \hat{t}_{i}\right)
\delta \left( \hat{t}-\hat{t}_{i}\right) ,  \label{1.4.11}
\end{equation}%
where $\delta \left( \bar{t}\right) $ is Dirac's delta-function, such that%
\begin{equation*}
\delta \left( \hat{t}\right) =\left\{ 
\begin{tabular}{l}
$0,\quad \hat{t}\neq 0;$ \\ 
$\infty ,\quad \hat{t}=0;$%
\end{tabular}%
\quad \int\limits_{-\infty }^{\infty }\delta \left( \eta \right) d\eta
=1.\right.
\end{equation*}

By analogy with (\ref{1.4.10}) and (\ref{1.4.11}), in (\ref{1.4.8}),%
\begin{equation*}
\tilde{Y}\left( \hat{t}\right) =\frac{1}{\nu }\left[ K_{0}+\sum_{i=0}^{n}%
\hat{I}\left( \hat{t}_{i}\right) \right] ,\quad 0\leq \hat{t}<\hat{t}_{n+1},
\end{equation*}%
\begin{equation*}
\frac{d\tilde{Y}\left( \hat{t}\right) }{d\hat{t}}=\frac{1}{\nu }%
\sum_{i=0}^{n}\hat{I}\left( \hat{t}_{i}\right) \delta \left( \hat{t}-\hat{t}%
_{i}\right) ,
\end{equation*}%
and the functions $\tilde{C}\left( \hat{t}\right) $, $\tilde{S}\left( \hat{t}%
\right) $, and $\tilde{I}\left( \hat{t}\right) $ in (\ref{1.4.5}) and (\ref%
{1.4.6}) have the same structure: in other words, all of them are
generalized functions. At that, the ordinary differential equation with
constant coefficients (\ref{1.4.8}) has, indeed, a solution of the form (\ref%
{1.1.8}) in the class of generalized functions \cite{3} (pp. 60, 61).

Can one, based on these arguments, draw a conclusion that%
\begin{equation}
\tilde{Y}\left( \hat{t}\right) =Y\left( \hat{t}\right) ,  \label{1.4.12}
\end{equation}%
where $Y\left( \hat{t}\right) $ is determined by expression (\ref{1.4.9}),
and that the model (\ref{1.1.1})-(\ref{1.1.5}) objectively reflects the
dynamics of the macroeconomic growth?

\section{Inadequacy of the initial model}

In the explanation to (\ref{1.1.5}), it is said: "the capital-to-national
income ratio is a constant quantity" \cite{1}. However, this fact
predetermines the solution of the problem. Indeed, for $\hat{t}=0,1,\ldots $%
, we have, respectively: $K_{0}$, $Y_{0}=K_{0}/\nu $;%
\begin{equation*}
I_{0}=K_{0}\mu /\nu ;\ K_{1}=K_{0}\left( 1+\alpha \right) ;\ \tilde{Y}%
_{1}=K_{1}/\nu ;\ \tilde{I}_{1}=K_{1}\mu /\nu ;\ \alpha =\mu /\nu ;
\end{equation*}%
\begin{equation*}
K_{2}=K_{0}\left[ 1+\left( 1+\alpha \right) \alpha \right] =K_{0}\left(
1+\alpha +\alpha ^{2}\right) ;\ \tilde{Y}_{2}=K_{2}/\nu ;\ \tilde{I}%
_{2}=K_{2}\mu /\nu ;\ldots ;
\end{equation*}%
\begin{equation}
K_{n}=K_{0}\sum\limits_{i=0}^{n}\alpha ^{i};\ \tilde{Y}_{n}=K_{n}/\nu ;\ 
\tilde{I}_{n}=K_{n}\mu /\nu .  \label{1.5.1}
\end{equation}%
As $\alpha <1$ (this follows from the content of the model, which is,
mathematically, insignificant), using the formula for a decreasing
geometrical progression, we get:%
\begin{equation}
\tilde{Y}_{n}=K_{0}\frac{1-\alpha ^{n+1}}{\nu \left( 1-\alpha \right) }=%
\tilde{Y}_{0}\frac{1-\alpha ^{n+1}}{1-\alpha }=\tilde{Y}_{0}\frac{1-\left(
\mu /\nu \right) ^{n+1}}{1-\mu /\nu },\ n=0,1,\ldots .  \label{1.5.2}
\end{equation}

Such a result contradicts the solution (\ref{1.4.9}) that, for $\hat{t}=\hat{%
t}_{n}=n$, has the form

\begin{equation}
\tilde{Y}_{n}=\tilde{Y}_{0}e^{\alpha n}=\tilde{Y}_{0}e^{\mu n/\nu },
\label{1.5.3}
\end{equation}%
because the equality of expressions (\ref{1.5.2}) and (\ref{1.5.3}) would
mean that%
\begin{equation}
e^{\alpha n}=\frac{1-\alpha ^{n+1}}{1-\alpha },\ \alpha n=\ln \frac{1-\alpha
^{n+1}}{1-\alpha }  \label{1.5.4}
\end{equation}%
for any $n=1,2,\ldots $ and $0<\alpha \,<1$, whereas the cases $n=0$ and $%
\alpha =0$ are trivial.

Using an analog of (\ref{1.5.4}) for $\hat{t}=\hat{t}_{n+1}$, we arrive at
the equation%
\begin{equation}
\alpha =\ln \frac{1-\alpha ^{n+1}}{1-\alpha ^{n}}  \label{1.5.5}
\end{equation}%
that is satisfied by $\alpha =1$. However, in this case, (\ref{1.5.1}) does
not represent any progression, and, instead of (\ref{1.5.2}) and (\ref{1.5.4}%
), we get:%
\begin{equation*}
\tilde{Y}_{n}=\left( n+1\right) \tilde{Y}_{0};\ e^{n}=n+1.
\end{equation*}

This equation cannot be satisfied for any $n=1,2,\ldots $, and it becomes
apparent that the values of $\tilde{Y}_{n}$ determined by formulas (\ref%
{1.5.2}) and (\ref{1.5.3}) are cardinally different. Here, one cannot assume
any approximation of the exponential dependence (\ref{1.5.3}) by means of
the rational representation (\ref{1.5.2}): see \cite{4} (pp. 19-24). As a
matter of fact, the same conclusion can be easily drawn by substituting
concrete values of $n$ and $\alpha $ into (\ref{1.5.5}).

However, a contradiction of the considered model is obvious:

- on the one hand, the function $Y\left( \hat{t}\right) $ determined by
expression (\ref{1.4.9}) satisfies equation (\ref{1.4.8});

- on the other hand, the same function in the form (\ref{1.5.3}) is rather
remote from (\ref{1.5.2}) that is, in essence, a programmed solution of the
same problem in the difference formulation.

Certainly, an explanation is related to the fact that both the functions $%
\tilde{Y}\left( \hat{t}\right) $ from section 1.4 and, accordingly, $Y\left(
t\right) $ in (\ref{1.1.7}) and (\ref{1.1.8}) can only be generalized
functions. In this situation, the equality (\ref{1.4.12}) is impossible in
principle, because, in reality, $\tilde{Y}\left( \hat{t}\right) $ is a
number appearing as a result of an operation on the function $Y\left( \hat{t}%
\right) $. In other words, there exists only a correspondence of the form%
\begin{equation}
\tilde{Y}\left( t\right) \thicksim \int\limits_{0}^{t_{n}}Y\left( \eta
\right) \varphi \left( \eta \right) d\eta ,  \label{1.5.6}
\end{equation}%
where $\varphi \left( \hat{t}\right) $ is an infinitely differentiable
functions on the interval $\left[ 0,\hat{t}_{n}\right] $ except for its
boundaries: it is called finite (alias trial or support) function.

For example,%
\begin{equation*}
\varphi \left( \hat{t}\right) =\left\{ 
\begin{tabular}{l}
$\sin \left( \pi \hat{t}/\hat{t}_{n}\right) ,\ \hat{t}\in \left[ 0,\hat{t}%
_{n}\right] ;$ \\ 
$0,\ 0>\hat{t}>\hat{t}_{n},$%
\end{tabular}%
\right.
\end{equation*}%
$\varphi \left( \hat{t}_{n}/2\right) =1$. Moreover, the derivative in (\ref%
{1.1.4}) is understood as follows \cite{3} (p. 34):%
\begin{equation*}
\int\limits_{0}^{t_{n}}\dot{K}\left( \eta \right) \varphi \left( \eta
\right) d\eta =-\int\limits_{0}^{t_{n}}K\left( \eta \right) \dot{\varphi}%
\left( \eta \right) d\eta .
\end{equation*}

As regards expressions (\ref{1.1.8}) and (\ref{1.4.9}), they can be
associated only with the function $\tilde{Y}\left( \hat{t}\right) $ from (%
\ref{1.5.6}), which by no means facilitates evaluation of $Y\left( \hat{t}%
\right) $ at the points $\hat{t}=\hat{t}_{i}$. "But how can one define the
integral of the product of a generalized function and a trial function, if
one cannot work with values of the function at separate points? The answer
is simple: in this case, one should define the integral axiomatically rather
than constructively" \cite{5} (pp. 10, 11). According to an alternative
theory \cite{6}, generalized functions are introduced as specially defined
limits of series of continuous functions. From the point of view of
reanimating the model (\ref{1.1.1})-(\ref{1.1.5}), they are also useless.

I. M. Gelfand and G. E. Shilov point out: "In the solution of concrete
problems of mathematical physics, the delta-function (as well as other
singular functions) appear, as a rule, only at Intermediate stages. In the
final answer, singular functions are either altogether absent or figure
under the sign of integration in a product with some sufficiently good
function. Thus, there is no direct need to answer the question what a
singular function is in itself. It is sufficient to answer the question what
the integral of a product of a singular function and a 'good' function is.
... In other words, we relate any singular function to a functional that
puts this singular function and any 'sufficiently good" function into
correspondence with a certain number" \cite{3} (pp. 14, 15).

However, why should we at all mention generalized functions, if their
techniques are useless in our case? The whole point is that, only for these
functions, the transformations leading to equation (\ref{1.1.7}) and,
accordingly, its solution are legitimate. The solution of (\ref{1.1.7}),
understood in the classical sense, is continuous: see, e.g., \cite{7}
(pp.28, 29). At the same time, relations (\ref{1.1.1})-(\ref{1.1.5}) are
fundamentally irresolvable in the class of continuous functions. One can
draw a conclusion that they are inadequate to the process of the
macroeconomic growth that is the subject of the model! What is more, the
following arguments prove to be completely unsuitable:

- macroeconomics operates the scales of decades;

- against this background, an income per year is only a small step;

- the function (\ref{1.1.8}) smooths out such steps, which reflects the
dynamics of growth on the whole.

However, we have here not only just "steps": as a matter of fact, they
induce a cardinal change of the type of the equation inherent in the
problem. The inadequacy of the problem is stipulated, in the first place, by
relation (\ref{1.1.5}) that provides it with the property of discreteness.
In this case, the function $Y\left( t\right) $ in (\ref{1.1.1})-(\ref{1.1.5}%
) embodies fundamentally different qualities: the intensity of the flow and
its scale on a finite time interval. The role of relation (\ref{1.1.5}) is
rather pragmatic: namely, as the derivative (\ref{1.1.4}) is objective, this
relation relates it in a proportional way to the function, which leads to
the differential equation (\ref{1.1.7}).

\chapter{Other models}

\section{Harrod-Domar's model}

In the work by R. Allen \cite{8} (pp. 75-78), this model and the resulting
dynamics of the growth are represented in a dimensionless form:%
\begin{equation}
\bar{Y}\left( \bar{t}\right) =\bar{C}\left( \bar{t}\right) +\bar{I}\left( 
\bar{t}\right) ;\ \bar{C}\left( \bar{t}\right) =\left( 1-\mu \right) \bar{Y}%
\left( \bar{t}\right) ;\ \bar{I}\left( \bar{t}\right) =\nu \frac{d\bar{Y}%
\left( \bar{t}\right) }{d\bar{t}},  \label{2.1.1}
\end{equation}%
where $0<\mu <1$; $\nu >0$. The author emphasizes that $\bar{Y}\left( \bar{t}%
\right) $, $\bar{C}\left( \bar{t}\right) $, and $\bar{I}\left( \bar{t}%
\right) $ are "functions of continuously changing time". The solution of the
problem is obtained in the form%
\begin{equation}
\bar{Y}\left( \bar{t}\right) =\bar{Y}_{0}e^{\mu \bar{t}/\nu }.  \label{2.1.2}
\end{equation}

For greater clearness, it is reasonable to transform (\ref{2.1.1}) and (\ref%
{2.1.2}) into dimensional notation:%
\begin{equation}
t=t_{0}\bar{t};\ Y\left( t\right) =Y_{0}\bar{Y}\left( \bar{t}\right) ;\
C\left( t\right) =C_{0}\bar{C}\left( \bar{t}\right) ;\ I\left( t\right)
=I_{0}\bar{I}\left( \bar{t}\right) ,  \label{2.1.3}
\end{equation}%
where $Y_{0}$, $C_{0}$, and $I_{0}$ are the intensities of corresponding
flows ($\$/s$).

We get:%
\begin{equation}
Y\left( t\right) =k_{1}C\left( t\right) +k_{2}I\left( t\right) ;\
k_{1}C\left( t\right) =\left( 1-\mu \right) Y\left( t\right) ;  \label{2.1.4}
\end{equation}%
\begin{equation}
k_{2}I\left( t\right) =\nu t_{0}\dot{Y}_{0},  \label{2.1.5}
\end{equation}%
where $k_{1}=Y_{0}/C_{0}$; $k_{2}=Y_{0}/I_{0}$, which leads to the
differential equation%
\begin{equation}
\nu t_{0}\dot{Y}\left( t\right) =\mu Y\left( t\right) ,  \label{2.1.6}
\end{equation}%
whose solution is%
\begin{equation}
Y\left( t\right) =Y_{0}e^{\mu t/\nu t_{0}}.  \label{2.1.7}
\end{equation}

However, let us draw attention to the fact that the confusion of
double-faced character of the function $Y\left( t\right) $ from the previous
section has disappeared, and the situation has become completely
transparent. Indeed, all the functions in (\ref{2.1.4}) are intensities of
the flows; one cannot raise any objections against the derivation of
equation (\ref{2.1.6}) and its solution (\ref{2.1.7}).

Nevertheless, \ expression (\ref{2.1.7}) constitutes a clear proof of the
inadequacy of this model. The main point is the dependence of $Y\left(
t\right) $ on the dimension of the parameter $t_{0}$ from (\ref{2.1.3})
that, by definition, is chosen arbitrarily. In other words, a mathematical
model, under the condition of its reliability, cannot depend on the scale of
time.

Accordingly, relations (\ref{2.1.4}) and (\ref{2.1.5}) do not contain
objective meaning. Generally speaking, the first two of them are quite
elementary, whereas (\ref{2.1.5}), the so-called accelerator, essentially
introduces a derivative. On can assume that, in this case, the problem was
solved the other way, namely, in relationship to obtaining differential
relations of the form (\ref{1.1.6}).

Note that the model (\ref{2.1.1}) does not employ the function $K\left(
t\right) $, the capital, whose special role is pointed out at the beginning
of section 1.1 with reference to \cite{1}. Indeed, the dependence (\ref%
{1.1.9}), as well as its corollary (\ref{1.1.4}), can be called fundamental.
The substitution of $I\left( t\right) $ from (\ref{1.1.4}) into (\ref{2.1.5}%
) yields:%
\begin{equation*}
k_{2}\dot{K}\left( t\right) =\nu t_{0}\dot{Y}_{0},
\end{equation*}%
from which we get%
\begin{equation}
K_{1}\left( t\right) =\left( \nu /k_{2}\right) t_{0}Y_{1}\left( t\right) ,
\label{2.1.8}
\end{equation}%
where%
\begin{equation*}
K_{1}\left( t\right) =K\left( t\right) -K_{0};\ Y_{1}\left( t\right)
=Y\left( t\right) -Y_{0}
\end{equation*}%
are, respectively, an increase of the capital during the period of time from 
$0$ and $t$ and the intensity of the income at the moment $t$.

How can the capital in (\ref{1.1.8}) be equal to the product (with a certain
coefficient) of the intensity of the income at the final moment of the term
of accumulation and an arbitrary period of time? However, relation (\ref%
{2.1.8}) is, practically, an analogue of (\ref{1.1.5}) if one assumes $%
t_{0}=t_{\ast }=1$ year. Accordingly, there appears discreteness (see
section 1.4) that in the same way leads to a conclusion that the considered
model is inadequate (see section 1.5)!

It should be noted that R. Allen himself does not touch on the issue of
dimension. It can be firmly grasped only on the basis of numerical examples.
In this sense, the situation with the model of section 1.1 was more
complicated. By definition \cite{8} (p. 193),%
\begin{equation}
\bar{I}\left( \bar{t}\right) =\frac{d\bar{K}\left( \bar{t}\right) }{d\bar{t}}%
.  \label{2.1.9}
\end{equation}%
Accordingly, for $\bar{K}\left( \bar{t}\right) =K\left( t\right) /K_{0}$, by
analogy with (\ref{2.1.3}), to reduce (\ref{2.1.9}) to the form (\ref{1.1.4}%
), it is necessary that $t_{0}=K_{0}/I_{0}$. In this case, however, instead
of the coefficient $\nu $, we get $\mu \nu ^{2}$ in (\ref{1.1.5}). If only
the variable $t$ is dimensionless, instead of (\ref{1.1.4}), we get a
relation of the type (\ref{1.4.3}). This list can be continued.

As is pointed out by T. Puu \cite{9} (p. 76), exactly R. Harrod is the
author of the idea of the formulation of the macroeconomic model in
continuous time by means of a differential equation (1948). At the same
time, an analysis of the basic work by P. Harrod \cite{10}, also published
in 1948, does not confirm this fact. Computational relations in this work
are given in a discrete form. On the basis of the same approach the model of
Harrod, as well as the models of Domar, of Solow, and of Samuelson and
Hicks, are considered by A. Lusse \cite{11} (pp. 163-166, 172-179, 180-182).
L. Stoleru is of the same opinion about Harrod's model \cite{12} (pp.
272-277). On the whole, one can assume that it is not R. Harrod to whom
dubious merit of inventing the above-discussed models should be attributed.

\section{Phillips' model and other models}

$K\left( \bar{t}\right) $ is also absent from the accelerator-multiplier
model of Phillips considered by R. Allen. We restrict ourselves to the
equations of the general solution \cite{8} (pp. 81, 82):%
\begin{equation}
\frac{d\bar{I}\left( \bar{t}\right) }{d\bar{t}}=-\kappa \left[ \bar{I}\left( 
\bar{t}\right) -\nu \frac{d\bar{Y}\left( \bar{t}\right) }{d\bar{t}}\right]
;\ \bar{Z}\left( \bar{t}\right) =\left( 1-\mu \right) \bar{Y}\left( \bar{t}%
\right) +\bar{I}\left( \bar{t}\right) ;  \label{2.2.1}
\end{equation}%
\begin{equation*}
\frac{d\bar{Y}\left( \bar{t}\right) }{d\bar{t}}=-\lambda \left[ \bar{Y}%
\left( \bar{t}\right) -\bar{Z}\left( \bar{t}\right) \right] ,
\end{equation*}%
where, in addition to the functions employed above, $Z\left( \bar{t}\right) $
is the intensity of cumulative demand. The coefficients are specified as
follows:

$\kappa $ is the rate of reaction, i.e., the inverse of a constant lag of
investment;

$\nu $ is the factor of the accelerator power;

$\mu $ is the multiplier;

$\lambda $ is the rate of the influence of the production (income) on the
demand. (Note that, in contrast to the model \cite{1} where in five
relations we had only one derivative, here, in only three relations we have
three derivatives.)

In the variables (\ref{2.1.3}), the system of equations (\ref{2.2.1}) takes
the form%
\begin{equation}
t_{0}\dot{I}\left( t\right) =-\kappa \left[ I\left( t\right) -\left( \nu
t_{0}/k_{2}\right) \dot{Y}\left( t\right) \right] ;\ k_{3}Z\left( t\right)
=\left( 1-\mu \right) Y\left( t\right) +k_{2}I\left( t\right) ;
\label{2.2.2}
\end{equation}%
\begin{equation*}
t_{0}\dot{Y}\left( t\right) =-\lambda \left[ Y\left( t\right) -k_{3}Z\left(
t\right) \right] ,
\end{equation*}%
where $Z\left( t\right) =Z_{0}\bar{Z}\left( \bar{t}\right) $; $%
k_{3}=Y_{0}/Z_{0}$.

Changing the variables according to (\ref{1.3.4}), we reduce the problem to
the solution of the second-order ordinary differential equation%
\begin{equation}
\ddot{Y}\left( t\right) +a\dot{Y}+bY\left( t\right) =0,  \label{2.2.3}
\end{equation}%
where $a=a_{1}/\rho $; $b=b_{1}/\rho ^{2}$,%
\begin{equation*}
a_{1}=\kappa +\mu \lambda -\kappa \nu \lambda ;\ b_{1}=\kappa \nu \lambda ,\
\rho =t_{0}/t_{\ast }.
\end{equation*}%
It has the form%
\begin{equation}
Y\left( \hat{t}\right) =c_{1}e^{p_{1}\hat{t}}+c_{2}e^{p_{2}\hat{t}},
\label{2.2.4}
\end{equation}%
where $c_{1}$ and $c_{2}$ are arbitrary constants;%
\begin{equation*}
p_{1,2}=\frac{1}{2\rho }\left[ a_{1}\pm \sqrt{a_{1}^{2}-4b_{1}}\right]
\end{equation*}%
are the roots of the quadratic equation $p^{2}+ap+b=0$.

As in section 2.1 the transformations of the flows are quite correct.
However, the situation is completely analogous to that considered in section
2.1. Because of the presence of the parameter $\rho =t_{0}/t_{\ast }$ in (%
\ref{2.2.4}), the solution depends on the choice of the scale of time, which
makes the model inadequate. Otherwise, with the help of $\rho $, we could a
priori set the periods of oscillations.

Using (\ref{1.1.4}), we express $I\left( t\right) $ in (\ref{2.2.2}) via $%
K\left( t\right) $:%
\begin{equation*}
t_{0}\ddot{K}\left( t\right) =-\kappa \left[ \dot{K}\left( t\right) -\left(
\nu t_{0}/k_{2}\right) \dot{Y}\left( t\right) \right] ;\ k_{3}Z\left(
t\right) =\left( 1-\mu \right) Y\left( t\right) +k_{2}\dot{K}\left( t\right)
;
\end{equation*}%
\begin{equation*}
t_{0}\dot{Y}\left( t\right) =-\lambda \left[ Y\left( t\right) -k_{3}Z\left(
t\right) \right] ,
\end{equation*}%
which yields%
\begin{equation*}
t_{0}\ddot{K}\left( t\right) =-\kappa \dot{K}\left( t\right) +\left( \kappa
\nu t_{0}/k_{2}\right) \dot{Y}\left( t\right) ;\ t_{0}\dot{Y}\left( t\right)
=-\mu \lambda Y\left( t\right) +\lambda k_{2}\dot{K}\left( t\right)
\end{equation*}%
As a result of the substitution%
\begin{equation*}
Y\left( t\right) =-\frac{k_{2}}{\kappa \nu \mu \lambda }\left[ t_{0}\ddot{K}%
\left( t\right) +\kappa \left( 1-\nu \lambda \right) \dot{K}\left( t\right) %
\right]
\end{equation*}%
and some simple transformations, we get%
\begin{equation}
t_{0}^{2}\dddot{K}\left( t\right) +t_{0}\left( \kappa +\mu \lambda -\kappa
\nu \lambda \right) \ddot{K}\left( t\right) +\kappa \nu \lambda \dot{K}%
\left( t\right) =0.  \label{2.2.5}
\end{equation}

Here, absolutely inappropriate dependence of the solution on the scale of
time is obvious. However, let us put aside this cardinal issue and assume
that $t_{0}=t_{\ast }$, as a law of Nature. It is not difficult to find that
the coefficients of equations (\ref{2.2.3}) and (\ref{2.2.5}) completely
coincide. At the same time, on the basis of (\ref{1.1.4}), equation (\ref%
{2.2.5}) is a second-order differential equation with respect to the
investment $I\left( t\right) $.

Accordingly, the functions $Y\left( t\right) $ and $I\left( t\right) $ may
differ from each other only in the constants $c_{1}$ and $c_{2}$ in the
representation of the solutions of the type (\ref{2.2.4}). To find them, we
need initial conditions%
\begin{equation*}
Y\left( 0\right) =Y_{0};\ \dot{Y}\left( 0\right) =\dot{Y}_{0};\ I\left(
0\right) =I_{0};\ \dot{I}\left( 0\right) =\dot{I}_{0}
\end{equation*}%
that are rather ephemeral. Combined with abstract character of the
coefficients of equations (\ref{2.2.1}), this fact makes the model
practically useless.

Besides, the considered model ignores, in fact, the function $K\left(
t\right) $, i.e., the capital, whose exclusiveness as a unique "restricted
factor" is emphasized by the authors of \cite{1}: see section 1.1. Indeed,
dynamic process, by virtue of objective circumstances, should be related to
a comparatively stable factor, because otherwise, figuratively speaking, a
reference point is lost.

Obviously, even under the condition of a "law of Nature", the model (\ref%
{2.2.1}) has no rehabilitation potential. In other words, it cannot be
converted even into a conditionally adequate one that would be analogous to
the transformations of Harrod-Domar's model considered in section 1.3.

In light of the above, let us turn to the same model of Phillips in the
interpretation of L. Bergstom \cite{13} (pp. 40, 41):%
\begin{equation*}
\bar{C}\left( \bar{t}\right) =\left( 1-\mu \right) \bar{Y}\left( \bar{t}%
\right) ;\ \frac{d\bar{Y}\left( \bar{t}\right) }{d\bar{t}}=\lambda \left[ 
\bar{C}\left( \bar{t}\right) +\frac{d\bar{K}\left( \bar{t}\right) }{d\bar{t}}%
-Y\left( \bar{t}\right) \right]
\end{equation*}%
\begin{equation}
\frac{d\bar{K}\left( \bar{t}\right) }{d\bar{t}}=\gamma \left[ \nu \bar{Y}%
\left( \bar{t}\right) -\bar{K}\left( \bar{t}\right) \right] ,  \label{2.2.6}
\end{equation}%
where $Y\left( \bar{t}\right) $ is a factual net income or output; $C\left( 
\bar{t}\right) $ is factual consumption; $K\left( \bar{t}\right) $ is the
volume of the capital; $\mu $, $\nu $, $\gamma $, and $\lambda $ are
positive constants ($\mu <1$).

It is explained: "As the capital represents the 'stock', whereas output is
an 'outflow', the quantity $\nu $ is inversely proportional to a chosen unit
of time. Thus, for example, when time in measured in months, $\nu $ is 12
times greater than in the case when time is measured in years."

The problem is reduced to the solution of the equation%
\begin{equation*}
\frac{d^{2}\bar{K}\left( \bar{t}\right) }{d\bar{t}^{2}}+\left( \gamma +\mu
\lambda -\nu \gamma \lambda \right) \frac{d\bar{K}\left( \bar{t}\right) }{d%
\bar{t}}+\mu \gamma \lambda \bar{K}\left( \bar{t}\right) =0
\end{equation*}%
[a comparison with (\ref{2.2.3}) yields $\gamma =\kappa $]. However, the
model contains the fundamental drawback that has been considered in detail
above. In other words, equation (\ref{2.2.6}) is nothing but an analogue of (%
\ref{1.1.5}). Taking into account (\ref{2.1.3}), for $K\left( t\right) =K_{0}%
\bar{K}\left( \bar{t}\right) $, we again arrive at the association of the
capital ($\$$) with the intensity of the flow $Y\left( t\right) $ ($\$/s$).
Nevertheless, as follows from the above-mentioned quotation, the author of 
\cite{13} fully admits a possibility of differentiating discrete flows.

Having worked out certain techniques, we can a priori identify the
components of macroeconomic models that contain contradictions. Thus, a
"simple" version of Goodwin's model \cite{8} (pp. 193, 194) involves the
equation%
\begin{equation*}
\bar{K}\left( \bar{t}\right) =\nu \bar{Y}\left( \bar{t}\right) +a\bar{t}:
\end{equation*}%
this situation is quite analogous to (\ref{2.2.6}).

From the point of view of the diagnosis, an "early" version of Kaletsky's
model \cite{8} (pp. 199-201), practically, does not differ much from the
above. It is based on the relation%
\begin{equation*}
\bar{I}\left( \bar{t}\right) =a\left( 1-c\right) \bar{Y}\left( \bar{t}%
\right) -k\bar{K}\left( \bar{t}\right) +\varepsilon ,
\end{equation*}%
where $a$ and $k$ are dimensionless constants, or, taking into account a lag,%
\begin{equation*}
\bar{Y}\left( \bar{t}\right) =\frac{1}{\theta \left( 1-c\right) }\left[ \bar{%
K}\left( \bar{t}+\theta \right) -\bar{K}\left( \bar{t}\right) \right] +\frac{%
A}{1-c}.
\end{equation*}

Its "later" version \cite{8} (pp. 205, 206), based on the equation%
\begin{equation*}
\bar{B}\left( \bar{t}\right) =a\left( 1-c\right) \bar{Y}\left( \bar{t}%
\right) +\nu _{2}\frac{d\bar{Y}\left( \bar{t}\right) }{d\bar{t}}-k\frac{d%
\bar{K}\left( \bar{t}\right) }{d\bar{t}}+\varepsilon ,
\end{equation*}%
where $\nu _{2}$ and $k$ are constants, mathematically, has practically the
same defects as (\ref{2.2.1}).

Phillips' multiplier model \cite{8} (p. 79)%
\begin{equation*}
\bar{Z}\left( \bar{t}\right) =\left( 1-\mu \right) \bar{Y}\left( \bar{t}%
\right) ;\ \frac{d\bar{Y}\left( \bar{t}\right) }{d\bar{t}}=-\lambda \left[ 
\bar{Y}\left( \bar{t}\right) -\bar{Z}\left( \bar{t}\right) \right]
\end{equation*}%
is directly subject to the criticism of section 2.1. Note that the model is
designed in such a way that it is impossible to access the function $\bar{K}%
\left( \bar{t}\right) $, i.e., the capital.

\section{Differential and difference models}

Of interest are the arguments of T. Puu: "Samuelson's invention (1939) of
the business cycle that combines the principles of the interaction between a
multiplier and of an accelerator, is, without any doubt, a tremendous event.
The fact that a combination of such simple reasons, i.e., the buyers'
expenditure of a certain part of their income on consumption and the
producers' preservation of a fixed relation between the capital and the
production volume induced cyclic changes, was simple, surprising and, at the
same time, convincing. This model has, so to say, scientific elegance. By
the way, it should be noted that its economic prerequisite was the
macroeconomic approach of Keynes.

The initial model had been proposed as a process with discrete time, that is
as a difference equation; later, it was rather skillfully developed in
detail by Hicks (1950). Harrod (1948) arrived at an idea to formulate this
process in continuous time, as a differential equation. The obtained system,
instead of generating cycles, caused a growth; nevertheless, he clearly
realized that the process of development was like balance on the edge of the
blade submerged in surrounding instability.

This established a tradition for several decades. Business cycles were
formed on the basis of difference equations, whereas development was formed
with the help of differential equations. Today, we could say that Samuelson
and Hicks chose by chance a second-order process, whereas Harrod decided in
favour of a first-order process. We could also admit that a choice between
discrete and continuous modelling does not change anything. Dynamic
processes of any order can be formulated in the form of difference equations
and in the form of differential equations.

A choice of the type of the model (discrete as compared with continuous) can
be regarded as a matter of pure convenience. For analytical purposes, when
we desire to apply theorems from vast literature on differential equations,
a continuous approach is preferable. However, when we want to apply this
model to an experimental time series that is inevitably discrete, we have to
use discretization" \cite{9} (pp. 76, 77).

A. Bergstrom's opinion about the freedom of choice of the model is
analogous: "One of the most important methodological problems of
constructing economic models is the question what equations should be used
to describe such models: differential or finite-difference equations.
Although many individual decisions are made in regular intervals of time
(say, once a week or once a month), variables observed by the econometrist
represent a result of many particular decisions made by different
individuals at different points of time. Moreover, the intervals of the
observation of most economic variables are considerably larger than the
intervals between decision-making represented by these variables. These
facts lead to the idea that variables of a typical economic model should be
regarded as continuous functions of time and that such a model should be
described by differential equations. ...

One more argument in favor of representing models in the form of
differential equations is that, even in the absence of continuous
observations of economic variables, the predictable continuous trajectories
of changes of these variables may prove to be of considerable value. Let us
assume, for instance, that, in the company's view, the volume of sales of
its products is closely related to the national income of the country. Then,
to forecast the sales, it is very useful to have a predicted continuous
trajectory of a change of the national income, although measurements of this
variable are carried out only once a year. A continuous model allows one to
obtain such a prediction using discrete measurements of economic variables
over the past period of time" \cite{13} (pp. 8-11).

The position of J. Casti is alternative: "In discrete time, the dynamics of
the system can be described by means of difference relations. The most
important property of such a description is that it gives us an idea of the
behavior of the system in a certain local neighborhood of the current state.
At that, it is implicitly assumed that local information can be somehow
'unified', which allows us to understand global (in time and space) behavior
of the system. Such an approach proved to be sufficiently justified for an
analysis of many physical and technical problems. However, the possibility
of its application in the case of less studied problems, especially of
systems of social-economic nature, is by no means obvious" \cite{14} (pp.
17, 18).

In addition, in light of the material of sections 1.1.-2.2, we have to refer
to the remark of R. Allen of somewhat ambiguous meaning: "It should be noted
that mathematical economics belongs to applied mathematics: it embodies a
union of mathematics and economics. Any in the least bit interesting results
in the field of mathematical economics can be provided only by an economist
that uses mathematical techniques" \cite{8} (p. 19).

\section{Abstracted model}

However, the above-quoted arguments \cite{13}, as well as partly \cite{9}%
,implicitly imply an approximation of the behavior of the considered system
that, in its turn, is a solution to an appropriate differential equation. In
other words, the mathematical model is not constructed in an immediate
relationship to an objective content of the process. This fact motivated the
choice of the title of the section.

As an example, we turn our attention to a simplified version of the model of
long waves \cite{15} (pp. 84, 85):%
\begin{equation*}
\dot{x}\left( t\right) =-p\left[ x\left( t\right) -qy\left( t\right) \right]
;\ \dot{y}\left( t\right) =-r\left[ y\left( t\right) -sz\left( t\right) %
\right] ;\ z\left( t\right) =x\left( t\right) -y\left( t\right) ,
\end{equation*}%
where $x\left( t\right) $ is the rate of an increase in labor productivity; $%
y\left( t\right) $ is the rate of an increase in capital endowment; $z\left(
t\right) $ is the rate of an increase in the profit rate; $p$, $q$, $r$, and 
$s$ are structural coefficients that can perform the following functions:

- adaptation of the model to the behavior of the observed system;

- estimation of related factors of the reliability of the model;

- forecasting of the behavior of the system (under the condition of adequacy
of the model).

In \cite{15}, it is pointed out: "The system of differential equations with
positive and negative feedbacks employed by the authors describes a process
analogous to a simple mechanical system such as, e.g., a pendulum, a spring,
elastic constructions with a damper, etc. However, there exists a
considerable difference between economic and mechanical systems. In the
simplest mechanics, elasticity and damping forces are usually independent of
the system, whereas in economics all the factors are mutually interrelated".

In particular, the following is established. Regular cycles appear for $s=-2$
and $p=r$. In the interval $p\in \left[ 0.10,0.12\right] $, periodic
undamped oscillations appear whose duration is 50-60 years. Cycles of twenty
years appear when $p$ and $r$ increase up to $0.34$. A more complete version
of the model was tested using the statistics of time series of the USA
during the period from 1989 to 1982. The following values of the "adaptation
coefficients" have been obtained: $p=0.048$ and $r=0.25$, which corresponds
to damped oscillations with the period of 53.7 years.

The considerable difference between the coefficients $p$ and $r$ is
objectively interpreted. However, the sensitivity of this model to a small
change in the coefficients is apparent. From this it follows that the model,
represented by a system of two first-order differential equations with
constant coefficients, has rather limited the potential of the description
of macroeconomic processes.

\section{Leontief's model}

In R. Allen's view, "the economist can learn much from the engineer: both
the ways of using mathematical methods and the ability to pose technical
problems" \cite{8} (p.19). He can be well understood: indeed, equation (\ref%
{2.2.3}) of Phillips' model contains four absolutely abstract coefficients $%
\kappa $, $\nu $, $\mu $, and $\lambda $. In addition, we have "independent
investment and consumption expenditures". They represent the free term of
the equation and, accordingly, should be given. It goes without saying, that
the fundamental defect caused by the dependence of the solution on a choice
of the scale of time is also present: see sections 2.1 and 2.2.

However, equation (\ref{2.2.3}) is widely used in technics. Thus, it
describes free oscillations of the mass suspended to a spring under the
condition of viscous drag. All the parameters of such a system are concrete,
they can be measured, and the same concerns the external force in the case
of forced oscillation. The differential equation is rigorously derived on
the basis of the laws of mechanics: see, e.g., \cite{16} (pp. 43-49).

So, what can be learnt from the engineer, if we aiming at developing an
analogous approach to mathematical modelling in macroeconomics? It seems
that anything like that is impossible by definition. In economics, there are
no laws for idealized objects that can be put into correspondence with a
material point. However, economics, in its turn, has great advantage over
mechanics that is embodied in the equation of the balance of financial flows!

Namely, Leontief's model is brilliant both with respect to its simplicity
and efficiency related to the development of computer technologies. For
functions that depend discretely on time, this model was studied by M.
Morishima. It was pointed out: "So-called Leontief's model, initially static
by its nature, is usually transformed to a dynamic one by the insertion of
consumption and output lags, of the lifetime of the means of production, of
accelerators, of the growth of final demand, of structural changes caused by
technological innovations, etc." \cite{17} (p. 77).

One can draw a conclusion that the possibilities of mathematical techniques
of mechanics and economics, figuratively speaking, have different vectors.
The main thing is their reasonable application rather than following the
principles of blind imitation of approaches from those fields of knowledge
that seem to be the most mathematized ones from a superficial point of view.
By the way, although the engineer is aware of exact solutions for a bar, a
plate, and other elements of this kind, he experiences considerable
difficulties when designing unique constructions. He has no restrictive
criterion for the whole set of elements. The balance is absent!

In light of the above, proceeding to the constructive part of our
consideration, we consider, following \cite{18} (pp. 40-45), the dynamic of
a system defined as%
\begin{equation}
X\left( t\right) =F\left( X\left( t-\tau \right) \right) ,  \label{2.5.1}
\end{equation}%
where $X\left( t\right) $ is a vector function of a set of flows $%
x_{i}\left( t\right) $; $\tau _{m}=\max \left\{ \tau _{i}\right\} $ is a
maximum delay (lag) during which a change of the state can be taken to be
linear.

The dynamics of the system is characterized by the derivative of the
function (\ref{2.5.1}):%
\begin{equation}
\dot{X}\left( t\right) =\frac{dF}{dX}\dot{X}\left( t-\tau \right) +\frac{dF}{%
dt}X\left( t-\tau \right) .  \label{2.5.2}
\end{equation}%
From here, taking into account that $\tau _{m}$ is small, we can get:%
\begin{equation}
\dot{X}\left( t\right) =\frac{1}{\tau _{m}}\frac{dF}{dX}\left[ X\left(
t\right) -X\left( t-\tau \right) \right] +\frac{dF}{dt}X\left( t-\tau
\right) ,  \label{2.5.3}
\end{equation}%
or%
\begin{equation}
\dot{X}\left( t\right) =G\left( t,X\left( t\right) ,X\left( t-\tau \right)
\right) .  \label{2.5.4}
\end{equation}

At the same time, because of the presence in (\ref{2.5.3}) of the small
factor $\tau _{m}$ by the derivative, equation (\ref{2.5.4}) belongs to the
class of singularly perturbed equations. Both the study of such equations
and their numerical realization require the use of techniques from a rather
special arsenal: see \cite{19} (sections 2, 7) and also \cite{20}, \cite{21}
(section 1).

In the next section, we present a method of the reduction of Leontief's
balance model to a differential form which is alternative to the use of (\ref%
{2.5.1}) and (\ref{2.5.2}) that lead to (\ref{2.5.4}). However, if the
function $F\left( X\right) $ in (\ref{2.5.1}) is linear, equation (\ref%
{2.5.3}) becomes meaningless by virtue of its triviality, and, accordingly,
the derivation of a differential equation proves to be fundamentally
impossible. In this regard, we note that the issue of limitations of a
linear theory has been indirectly touched on also in section 2.4. This fact
is related to the content of section 3.2.

\chapter{Constructive arguments}

\section{Leontief's differential model}

Consider a system of balance equations:%
\begin{equation}
x_{i}\left( t\right) =\sum_{j=1}^{n}a_{ij}x_{j}\left( t\right) +c_{i}\left(
t\right) ,\ i=1,2,\ldots ,n;\ 0\leq t\leq t_{0}  \label{3.1.1}
\end{equation}%
that can be interpreted, e.g., as follows: $x_{i}\left( t\right) $ are the
flows of the volumes of output ($\$/s$); $a_{ij}$ is a part of the commodity 
$i$ used in the production of the commodity $j$; $c_{i}\left( t\right) >0$
is the flow of an external demand for the commodity $i$; $t$ is the variable
of time ($s$). In what follows, the statement of the problem will be
complemented. In matrix notation, the system of equations (\ref{3.1.1})
takes the form%
\begin{equation}
X\left( t\right) =AX\left( t\right) +C\left( t\right) ,\ 0\leq t\leq t_{0}.
\label{3.1.2}
\end{equation}

In this case, it is reasonable to assume that the matrix $A$ satisfies the
conditions of Metzler's theorem \cite{17} (p. 34), and, accordingly,%
\begin{equation*}
\sum_{j=1}^{n}a_{ij}\leq 1,\ i=1,2,\ldots ,n,
\end{equation*}%
where we have a strict inequality for at least one of the sums. Then,
equation (\ref{3.1.2}) can be solved by the method of simple iterations \cite%
{22} (pp. 120-122):%
\begin{equation}
X_{s+1}\left( t\right) =AX_{s}\left( t\right) +C\left( t\right) ,\
s=0,1,\ldots ,  \label{3.1.3}
\end{equation}%
where, in theory, the initial element can be chosen arbitrarily.

However, we have the initial condition $X\left( 0\right) =X_{0}$, and,
accordingly, the iterations (\ref{3.1.3}) have, in reality, the form%
\begin{equation}
X\left( t_{s+1}\right) =AX\left( t_{s}\right) +C\left( t_{s}\right) ,
\label{3.1.4}
\end{equation}%
where $t_{s+1}=t_{s}+\Delta t$, and there emerges the issue of choosing the
step $\Delta t$ that corresponds to the factors of changes in the vector
function $X\left( t\right) $.

To approximate this function, one employs a Taylor series expansion and
usually retains the first-order derivative \cite{23} (pp. 32, 33). However,
if we only want to capitalize on the idea of the outlined approach and,
instead of (\ref{3.1.4}), consider the relation%
\begin{equation}
X\left( t+t_{0}\right) =AX\left( t\right) +C\left( t\right) ,\ 0\leq t\leq
t_{0},  \label{3.1.5}
\end{equation}%
where the time interval $t_{0}$ is sufficiently large, we will have to
retain in the Taylor series%
\begin{equation}
X\left( t+t_{0}\right) =X\left( t\right) +t_{0}\dot{X}\left( t\right) +\frac{%
1}{2}t_{0}^{2}\ddot{X}\left( t\right) +\frac{1}{6}t_{0}^{3}\dddot{X}\left(
t\right) \ \ldots  \label{3.1.6}
\end{equation}%
more terms.

If, as an illustration, we restrict ourselves to three terms and use the
notation $A=E-B$, where%
\begin{equation}
B=\left( 
\begin{array}{cccc}
1-a_{11} & a_{12} & \ldots & a_{1n} \\ 
a_{21} & 1-a_{22} & \ldots & a_{2n} \\ 
. & . & \ldots & . \\ 
a_{n1} & a_{n2} & \ldots & 1-a_{nn}%
\end{array}%
\right) ,  \label{3.1.7}
\end{equation}%
and%
\begin{equation*}
E=\left( 
\begin{array}{cccc}
1 & 0 & \ldots & 0 \\ 
0 & 1 & \ldots & 0 \\ 
. & . & \ldots & . \\ 
0 & 0 & \ldots & 1%
\end{array}%
\right)
\end{equation*}%
is a unit matrix, after substitution into (\ref{3.1.5}), we obtain%
\begin{equation}
t_{0}\dot{X}\left( t\right) +0.5t_{0}^{2}\ddot{X}\left( t\right) =-BX\left(
t\right) +C\left( t\right)  \label{3.1.8}
\end{equation}%
that under a change of the variable $\bar{t}=t/t_{0}$ becomes%
\begin{equation}
0.5\ddot{X}\left( \bar{t}\right) +\dot{X}\left( \bar{t}\right) +BX\left( 
\bar{t}\right) =C\left( \bar{t}\right) ,\ 0\leq \bar{t}\leq 1.  \label{3.1.9}
\end{equation}

As a matter of fact, a derivation of this equation has been the aim of the
previous transformations.\ In the solution of this equation, one uses
initial condition of the form%
\begin{equation}
X\left( 0\right) =X_{0},\ \dot{X}\left( 0\right) =\dot{X}_{0},
\label{3.1.10}
\end{equation}%
and, despite its matrix form, the procedure of numerical realization is
quite standard \cite{24} (section 10). This reference also gives the
solution by quadrature of the first-order differential equation with the
elements of the matrix (\ref{3.1.7}) depending on time, i.e., for $%
a_{ij}=a_{ij}\left( t\right) $. In principle, this result can be easily
extended to the case of the second order as well as of higher order: see,
e.g., \cite{7} (section 5).

However, one may ask whether we are resorting to a double standard here
because the procedure of the reduction of (\ref{3.1.8}) and (\ref{2.2.2}) to
the forms (\ref{3.1.9}) and (\ref{2.2.3}), respectively, is the same.
Moreover, why the dependence of the solution on the time scale has not been
pointed out in this case? This is a very important issue resulting from the
use of the Taylor series (\ref{3.1.6}). The number of retained terms in this
series directly depends on the value of $t_{0}$.

The difference lies in the fact how this dependence is established: by a
trial-and-error method or by a profound analysis. As regards the
consideration of sections 2.1 and 2.2, there was no criterion of this kind
there.

In the formulation (\ref{3.1.1}), as the function $C\left( \bar{t}\right) $
is not given, the demand is prevailing and, in practice, the output may fail
to meet it. To avoid this situation, we can multiply $C\left( \bar{t}\right) 
$ by a factor $0<\alpha \,<1$ defined, e.g., by the condition%
\begin{equation*}
\int\limits_{0}^{1}X\left( \eta \right) d\eta =X_{\ast },
\end{equation*}%
where $X_{\ast }$ characterizes the demand for internal output. To carry out
analytical studies for forecast purposes, one can complement the vector $%
C\left( \bar{t}\right) $ by a component that nonlinearly depends on $X\left( 
\bar{t}\right) $.

\section{Nonlinear model and an alternative}

"The author has arrived at a final conclusion that linear dynamic modelling
yields very little. This is caused by the scarcity of the set of
alternatives, i.e., either damped or explosive motion, that are related to
linear models. Therefore, in the present study, we focus mainly on
nonlinearity" \cite{9} (p. 7). In what follows, we show that nonlinearity
has its own alternatives.

In our view, the work of A. van der Schaft \cite{25} is of great interest to
mathematical modelling. It contains information about the orientation of
system theory concerned with the realization of nonlinear dynamic models.
Thus realization implies the construction of a system of equations%
\begin{equation}
\dot{x}=f\left( x,u\right) ,\ x\left( 0\right) =x_{0},\ y=h\left( x,u\right)
,  \label{3.2.1}
\end{equation}%
where $x\left( t\right) $, $y\left( t\right) $, and $u\left( t\right) $ are
vector functions, that is adequate to a given input-output mapping%
\begin{equation}
y\left( t\right) =F\left( u\left( \tau \right) ;0\leq \tau \leq t\right) .
\label{3.2.2}
\end{equation}%
In other words, corresponding measurements can be carried out with a shift
in time, which is rather important. All the quantities are dimensionless.

As is pointed out by J. Casti, the problem of realization consists in the
construction, as far as possible, of a compact model that stands in
agreement with observed data \cite{14}.

Note that, for analytical studies of the nonlinear model (\ref{3.2.1}),.
there exist efficient mathematical techniques dating back to the works of A.
M. Lyapunov and A. Poincar\'{e}. The monograph \cite{26} is devoted to the
adaptation of these techniques to the problems of macroeconomics.

The author of \cite{25} has formulated an original part of his work in such
a way: "We consider systems of smooth nonlinear differential and algebraic
equations where certain variables are singled out as 'external' ones. The
problem of realization amounts to the substitution of an implicit high-order
differential equation by explicit first-order differential equations and
algebraic equations by means of a mapping in terms of external variables".

Thus, instead of (\ref{3.2.2}), a relationship between the vectors at the
input and output of the system, i.e., $u\left( t\right) $ and $y\left(
t\right) $, respectively, is implied in the form of a nonlinear high-order
differential equation. Certainly, with regard to the problem we are
considering, such formulation is absolutely irrelevant.

Indeed, in the course of consideration, we have been trying to derive a
differential equation of at least the first order, and we have been
discussing the correctness of the application of this procedure in the works
on macroeconomics that are considered to be classic. At the same time, the
mapping (\ref{3.2.2}) is rather organically related to the factors of the
balance of financial flows, to the realia of corresponding measurements as
well as to other factors of this kind. It may seem that one can be satisfied
with this fact. Why have we then referred to the above quotation?

The reason is that the text \cite{25} implies an alternative way of the
construction of macroeconomic models. Indeed, as a prerequisite of further
transformations, here appears an input-output mapping that is represented by
a system of linear differential equations%
\begin{equation}
D\left( \frac{d}{dt}\right) y\left( t\right) =N\left( \frac{d}{dt}\right)
u\left( t\right) ,  \label{3.2.3}
\end{equation}%
where $D\left( s\right) $ and $N\left( s\right) $ are polynomial matrices of
"appropriate dimensions", subject to a number of loose requirements. This
case is matched by the model (\ref{3.2.1}) under a linear approximation:%
\begin{equation*}
\dot{x}=Ax+Bu;\ Y=Cx+Du,
\end{equation*}%
where $A$, $B$, and $C$ are corresponding matrices.

Consider, however, an ordinary differential equation, with constant
coefficients as in (\ref{3.2.3}),$\ $\ of order $n$:%
\begin{equation}
z^{\left( n\right) }+a_{n-1}z^{\left( n-1\right) }+\ldots +a_{1}z=0,
\label{3.2.4}
\end{equation}%
where $z^{\left( n\right) }=\left( d/dt\right) ^{n}z\left( t\right) $.

Using a formal notation%
\begin{equation*}
z^{\left( n\right) }\left( t\right) =\varphi \left( t\right) ,
\end{equation*}%
we get%
\begin{equation}
z\left( t\right) =\frac{1}{\left( n-1\right) !}\int\limits_{0}^{t}\left(
t-\eta \right) ^{n-1}\varphi \left( \eta \right) d\eta +\sum_{i=1}^{n}c_{i-1}%
\frac{t^{i-1}}{\left( i-1\right) !}.  \label{3.2.5}
\end{equation}%
Here, $c_{i}$ are the constants of integration defined by conditions at the
input-output of the system, that is, at $t=0$ and, for definiteness, at $t=1$%
.

The substitution of the function (\ref{3.2.5}) and of its derivatives into (%
\ref{3.2.4}) leads to a Fredholm integral equation of the second kind in the
canonical form:%
\begin{equation}
\varphi \left( t\right) =\lambda \int\limits_{0}^{1}k\left( t,\eta \right)
\varphi \left( \eta \right) d\eta +q\left( t\right) ,\ t\in \left[ 0,1\right]
,  \label{3.2.6}
\end{equation}%
where $\lambda $ is a parameter discussed below. As a matter of fact, we
have described a classic method of the reduction of an ordinary differential
equation to an integral equation that is given in numerous publications:
see, e.g., \cite{27} (pp. 31-33).

Note that the use in (\ref{3.1.8}) of the notation%
\begin{equation*}
U\left( \bar{t}\right) =\ddot{X}\left( \bar{t}\right)
\end{equation*}%
yields%
\begin{equation*}
X\left( \bar{t}\right) =\int\limits_{0}^{\bar{t}}\left( \bar{t}-\eta \right)
U\left( \eta \right) d\eta +c_{0}\bar{t}+c_{1},
\end{equation*}%
which, under conditions (\ref{3.1.10}), results in Leontief's model in the
form of a Volterra integral equation of the second kind.

However, in light of the present consideration, the following point is of
primary importance. The whole information about the system, i.e., about
equation (\ref{3.2.4}) and conditions at $t=0$ and $t=1$, is reflected by
the kernel $k\left( t,\eta \right) $ and the free term $q\left( t\right) $
of equation (\ref{3.2.6}). At the same time, as a result of the repeated
integration according to (\ref{3.2.5}), the kernel in (\ref{3.2.6})
represents, in reality, a rather particular case, i.e., a convolution%
\begin{equation*}
k\left( t,\eta \right) =k\left( t-\eta \right) ,
\end{equation*}%
and this convolution is of a very specific type.

In this regard, one may ask a natural question: Based on a priori
information, measurements, experiments as well as other factors including
heuristics, why not construct the kernel $k\left( t,\eta \right) $ in such a
way that the function $\varphi \left( t\right) $ could possess the property
to represent potentially realizable variants of the behavior of the
considered system?

Simultaneously, the kernel $k\left( t,\eta \right) $ should contain
parameters intended for the purposes of adaptation and correction of the
model, as well as of forecasting. They can be associated with a part of
conditions at $t=1$ for the function $z\left( t\right) $ and its derivatives
from (\ref{3.2.4}) that are subject to identification. Without any doubt,
such a model contains elements of self-training.

However, could a model based on the integral equation (\ref{3.2.6}) prove to
be more efficient than (\ref{3.2.1}), that is, of the model posed in the
form of the Cauchy problem that used to be a traditional orientation of the
macroeconomic science?

From this point of view, E. Goursat's constatation is of interest: "The
solution of Volterra's integral equation is a broad generalization of the
Cauchy problem for a linear differential equation" \cite{28} (p. 16). At the
same time, however, equations of Volterra's type represent a particular case
of Fredhom equations whose spectrum of solutions is more diverse.

To answer the above-posed question, we consider the representativeness, in a
mathematical and, accordingly, objective sense, of the solution of equation (%
\ref{3.2.6}) that depends on the parameters of the model. In this regard,
solutions of the exponential type that were present in sections 1 and 2 can
be called trivial.

\section{Representativeness of the integral model}

Note that, for every set of the coefficients $a_{i}$ from (\ref{3.2.4}), the
parameter $\lambda $ in equation (\ref{3.2.6}) is a definite number. At the
same time, to study the properties of the function $\varphi \left( t\right) $
that satisfies (\ref{3.2.6}), it is reasonable to consider $\lambda $
indefinite and to specify it when necessary.

The solution to equation (\ref{3.2.6}) in the case of a sufficiently small
absolute value of the parameter $\lambda $ can be expressed via the
resolvent:%
\begin{equation*}
\varphi \left( t\right) =q\left( t\right) +\lambda
\int\limits_{0}^{1}H\left( t,\eta ,\lambda \right) q\left( \eta \right)
d\eta ,
\end{equation*}%
under the condition $\lambda \neq \lambda _{i}$, where $\lambda _{i}$ are
the characteristic numbers (alias points of the spectrum) of the kernel $%
k\left( t,\eta \right) $. The above-mentioned spectrum, together with $%
\lambda _{i}$ is associated with the eigenfunctions $\varphi _{i}\left(
t\right) $ such that%
\begin{equation}
\varphi _{i}\left( t\right) =\lambda _{i}\int\limits_{0}^{1}k\left( t,\eta
\right) \varphi _{i}\left( \eta \right) d\eta .  \label{3.3.1}
\end{equation}

Numerical realization of the resolvent $H\left( t,\eta ,\lambda \right) $ is
a special topic. On the whole, there exist a number of efficient algorithms
for the solution of Fredholm integral equations of the second kind. In any
case, both the parameter $\lambda $ and the function $H\left( t,\eta
,\lambda \right) $ are very convenient for analytical purposes. There are
publications devoted to the study of the structure of the resolvent in the
neighborhood of the characteristic numbers. Analytical estimates of the
order of the growth of the characteristic numbers depending on the
properties of the kernel are worked out \cite{29} (pp. 61, 62).

The following example clearly demonstrates the influence of the parameter $%
\lambda $ on the solution. Thus, the equation%
\begin{equation*}
\varphi \left( t\right) =\lambda \int\limits_{0}^{1}\left( t+\eta \right)
\varphi \left( \eta \right) d\eta +q\left( t\right) ,
\end{equation*}%
for $\lambda \neq -6\pm 4\sqrt{3}$, has the unique solution \cite{30} (pp.
29, 30)%
\begin{equation*}
\varphi \left( t\right) =q\left( t\right) +\frac{\lambda }{\lambda
^{2}+12\lambda -12}\int\limits_{0}^{1}\left[ 6\left( \lambda -2\right)
\left( t+\eta \right) -12t\eta -4\lambda \right] q\left( \eta \right) d\eta .
\end{equation*}

Equation%
\begin{equation}
\varphi \left( t\right) =\lambda \int\limits_{0}^{1}e^{t-\eta }\varphi
\left( \eta \right) d\eta +q\left( t\right) ,  \label{3.3.2}
\end{equation}%
for $\lambda \neq 1$, has the solution%
\begin{equation*}
\varphi \left( t\right) =q\left( t\right) +\frac{\lambda }{1-\lambda }%
e^{t}\int e^{-t}q\left( t\right) dt;
\end{equation*}%
if, however, $\lambda =1$ and $\int e^{-t}q\left( t\right) dt=0$, then%
\begin{equation*}
\varphi \left( t\right) =q\left( t\right) +ce^{t},
\end{equation*}%
where $c$ is an arbitrary constant. For $\lambda =\lambda _{i}$, equation (%
\ref{3.3.2}) is, in general, unsolvable, whereas the corresponding
homogeneous equation has an infinite number of nontrivial solutions \cite{27}
(pp. 74-76).

As is pointed out by V. I. Smirnov \cite{31} (p. 130), when considering
integral equations whose kernels are analytical functions of the parameter,
one can meet with substantial deviations from the regularities present in
general theory of Fredholm integral equations of the second kind. From this
point of view, of interest is the paper by Z. I. Khalilov \cite{32} devoted
to investigation into the equation%
\begin{equation}
\varphi \left( t\right) =\int\limits_{0}^{1}\left[ k_{0}\left( t,\eta
\right) +\mu k_{1}\left( t,\eta \right) \right] \varphi \left( \eta \right)
d\eta +q\left( t\right) ,  \label{3.3.3}
\end{equation}%
where $\mu $ is a parameter, under certain restrictions on $k_{0}\left(
t,\eta \right) $, $k_{1}\left( t,\eta \right) $, and $q\left( t\right) $.

Tamarkin's theorem is given: Equation (\ref{3.3.3}) either has no solutions
for all $q\left( t\right) $ and any value of $\mu $, or it has a unique
solution for all $q\left( t\right) $ and any $\mu $ except for, may be, a
countable set of values of $\mu $ together with a limit point (if it exists)
at the infinity. The statement of this theorem for a more general
formulation of $k\left( t,\eta ,\mu \right) $ is given in \cite{31} (p. 132).

In the first case, the author \cite{32} has termed the kernel of equation (%
\ref{3.3.3}) exceptional, whereas in the second case, he has termed it
non-exceptional. A theorem is proved stating that, in the case of a
non-exceptional kernel, equation (\ref{3.3.3}) can always be reduced to a
usual integral equation. Representations of the kernel of this equation $%
r\left( t,\eta \right) $ and of the free term via the functions and the
parameter from (\ref{3.3.3}), including the resolvent, are given.

One more theorem: The spectrum of the kernel in (\ref{3.3.3}) either
coincides with the whole plane of the complex variable $\mu $ or coincides
with the spectrum of the kernel $r\left( t,\eta \right) $. Conditions of the
solvability of equation (\ref{3.3.3}) in the case of multiple eigenvalues
are also obtained.

In the context of this consideration, the material presented by M. L.
Krasnov \cite{33} is relevant to the case. Thus, it turns out that the
homogeneous integral equation%
\begin{equation}
\varphi \left( t\right) =\int\limits_{0}^{1}\left[ \rho \left( t\right) \rho
\left( \eta \right) +\mu \sigma \left( t\right) \rho \left( \eta \right) %
\right] \varphi \left( \eta \right) d\eta ,  \label{3.3.4}
\end{equation}%
where $\rho \left( t\right) $ and $\sigma \left( t\right) $ are continuous
functions obeying the relations%
\begin{equation*}
\int\limits_{0}^{1}\rho ^{2}\left( \eta \right) d\eta =1;\
\int\limits_{0}^{1}\rho \left( \eta \right) \sigma \left( \eta \right) d\eta
=0,
\end{equation*}%
has the nonzero solution%
\begin{equation*}
\varphi \left( t\right) =\rho \left( t\right) +\mu \sigma \left( t\right)
\end{equation*}%
for any $\mu $ \cite{33} (p. 272). In other words, equation (\ref{3.3.4})
does not represent a Fredholm integral equation of the second kind.

F. Tricomi summarizes: "In the theory of equations that do not obey
Fredhom's theorems, three new phenomena often occur:

- the presence of finite limit points of the spectrum of characteristic
numbers or even of a continuous spectrum, i.e., of characteristic numbers
that fill the whole interval of the $\lambda $-axis and even the whole $%
\lambda $-axis;

- the presence of characteristic numbers of infinite multiplicity, i.e., of
characteristic numbers associated with an infinite number of linearly
independent functions;

- the presence of bifurcation points (in the real nonlinear case), i.e., of
those points of the $\lambda $-axis by passing through which the number of
solutions of the equation changes while remaining finite" \cite{27} (pp.
207, 208).

Note that the above-mentioned nonlinearity is not necessary for the
appearance of various miracles in the behavior of the solutions. As a matter
of fact, we initially have equation (\ref{3.2.6}) with the function $\varphi
\left( t\right) $ that is explicitly present. Exactly this type of equations
is necessary for the study of peculiarities of the solutions, such as
branching and bifurcation (they can be interpreted in different ways).

In the general case, to obtain an equation of the second kind, by analogy
with (\ref{2.5.1},) one needs to have a nonlinear initial object \cite{34}
(pp. 22, 23, 304-305, 318). By definition \cite{33}, the values of $\lambda
_{i}$ from (\ref{3.3.1}) representing characteristic numbers of the
continuous kernel $k\left( t,\eta \right) $ are also points of bifurcation
of equation (\ref{3.2.6}).

However, only rather special cases of the dependence of the kernel on the
parameter have been considered. Thus, Tamarkin's theorem does not envisage
the presence, alongside with $\mu $, of the parameter $\lambda $. At the
same time, if one is motivated by the development of an objective basis of
analytical investigation, it is desirable to make the number of the
parameters in the kernel sufficiently large. From a formal point of view,
there are no obstacles on this way. The main question is whether adequate
computational and theoretical means exist.

On the basis of a research into literature, we can arrive at a conclusion
that during the period whose beginning dates back to the publication \cite%
{35} there has been no more or less significant progress in the field of the
theory of integral equations with parameter-dependent kernels. How this fact
can be explained?

It seems that the following answer can be given. On the one hand, there are,
obviously, considerable difficulties of fundamental character on the way of
the development of mathematically rigorous theory. On the other hand, the
theory of integral equations has become part of functional analysis, and, at
the modern stage, its problems are considered from a more abstract point of
view (the theory of an implicit function). In a certain sense, the factor of
nonlinearity may prove to be of minor importance here \cite{34} (section 1).

In light of the above, we think that it would be rather interesting to
develop the theory of Volterra integral equations of the second kind with
the parameter-dependent kernel. On the one hand, the theory of equations of
this type is objectively simpler that of Fredholm. On the other hand, such
equations are associated in a natural way with the procedure of the
reduction of Leontief's model to the integral form: see sections 3.1 and 3.2.

Remaining within the framework of the objective orientation of the present
consideration, we can draw the following conclusions:

- the solution to an integral equation of the second kind even in a rather
particular case of dependence of the kernel on the parameter exhibits a wide
range of possibilities of behavior;

- as is illustrated by the simplest examples, the parameter can put the
equation outside the class of Fredholm equations of the second class that
are characterized by stable dependence of the solution on the data of the
problem;

- such properties of the seemingly linear integral equation of the second
kind with the parameter-dependent kernel characterize it as a rather
efficient tool for modelling of dynamic processes;

- at the same time, the breadth of the class of solutions results in
difficulties of an analytical study and of numerical realization of
mathematical models even in the case of dependence on the parameter of a
particular type.

\newpage

\begin{center}
\textbf{SUMMARY}

\bigskip
\end{center}

Traditional models of macroeconomic dynamics are fundamentally incorrect.
The reason lies in a misunderstanding of peculiarities of the analysis of
infinitesimal quantities. However, even those types of solutions that are
envisaged by the above-mentioned models are nonrepresentative in the sense
of the reflection of realities. It became obvious that the techniques of the
theory of linear differential equations were insufficient here. Accordingly,
the scientists' attention switched to the theory of nonlinear differential
equations.

At the same time, balance and, accordingly, the model with matrix properties
are objectively inherent in the economic system. For the reduction of this
model to a differential form, there exist rather elementary means that
proved to be unclaimed. Macroeconomic rhetoric - the power of the
accelerator, a lag on the part of demand, etc. - accompanied by the use of a
lot of abstract coefficients prevailed.

Why such an entourage? One cannot but get an impression that the issue, in
essence, is political and is deeply rooted. Economics is seemingly a major
science because it is more intimately related to the capital than other
sciences and it works out wise recommendations for the ruling elite.
Therefore, economics is prescribed to use higher mathematics whose peak is
differential equations: this has been a mentality since long ago. Nobody has
heard anything about other equations, whereas operating the categories of
balance is inappropriate because it looks like school arithmetic. The elite,
in its turn, acts according to prescriptions of science rather than at will.

However, there is no organic interrelation between matrix and nonlinear
differential equations. On the contrary, it can be said that linear theory
of integral equations originated in matrix analysis. The Fredholm linear
integral equation of the second kind with a parameter-dependent kernel
proves to be rather representative with regard to the class of possible
solutions. It seems that it can be used for the description of any zigzags
of the economy. The price one has to pay for this is the nontriviality of
existing theory.

\bigskip

\end{document}